\documentclass{v15cosmo}

\usepackage{float}
\usepackage{subfig}
\usepackage{amsmath}

\def\be{\begin{equation}}
\def\ee{\end{equation}}
\def\bea{\begin{eqnarray}}
\def\eea{\end{eqnarray}}

\newcommand{\Photo}{}

\begin{document}
\vspace*{4cm}
\title{Recovering hidden signals of statistical anisotropy\\from a masked or partial CMB sky}

\author{Pavan K. Aluri$^1$, Nidhi Pant$^1$, Aditya Rotti$^2$, Tarun Souradeep$^1$}

\address{$^1$Inter-University Centre for Astronomy and Astrophysics, Pune 411007, India\\
         $^2$Department of Physics, Florida State University, Tallahassee, FL 32304, USA}

\maketitle\abstracts{
Any isotropy violating phenomena on cosmic microwave background (CMB) induces off-diagonal correlations
in the two-point function. These correlations themselves can be used to estimate the underlying anisotropic signals.
Masking due to residual foregrounds, or availability of partial sky due to survey limitation, are unavoidable circumstances in CMB studies. But, masking induces additional correlations, and thus complicates the recovery of such signals. In this work, we discuss a procedure based on bipolar spherical harmonic (BipoSH) formalism to comprehensively addresses any spurious correlations induced by masking and successfully recover hidden signals of anisotropy in observed CMB maps. This method is generic, and can be applied to recover a variety of isotropy violating phenomena. Here, we illustrate the procedure by recovering the subtle Doppler boost signal from simulated boosted CMB skies, which has become possible with the unprecedented full-sky sensitivity of PLANCK probe.
}

\section{Motivation}
In the standard model of cosmology, the cosmic microwave background (CMB)
which is a prediction of the Big Bang theory of cosmic origin is expected
to be isotropic. This is so because of the so called
\emph{Cosmological Principle} which states that the universe is homogeneous
and isotropic on very large scales. Indeed CMB is highly isotropic to approximately
1 in $10^5$ parts validating the cosmological principle.
Further the CMB anisotropies are also expected to be statistically isotropic.
Yet there exists hints for deviations from statistical isotropy in CMB
data~\cite{wmap7yranom,plk13is,plk15is}.

A cleaning procedure is employed on the raw satellite data
to obtain a \emph{clean} map of the cosmic CMB signal that is of our prime interest. However in practice
any cleaning procedure leaves (albeit readily visible) foreground residuals in
regions of significant foreground emission, particularly in the galactic plane.
The recovered CMB signal in these regions is unreliable and needs to be omitted/masked to obtain
unbiased estimates of quantities of our interest such as power spectrum or
cosmological parameters. Likewise tests of statistical isotropy (SI) violation,
and any claims there off, have to be robust against such residuals which can bias
our inferences, if any deviations were to be found. But masking a CMB map introduces
additional correlations between various modes, and jeopardizes a reliable test/estimate
of any underlying signal of isotropy breakdown in a clean CMB map.

Here we present an estimator based on bipolar spherical harmonic (BipoSH) formalism,
that comprehensively addresses effects of masking, and facilitates a reliable test or
recovery of such hidden SI violating signals.

A CMB anisotropy field, $T(\hat{n})$, defined on a sphere is conventionally decomposed in terms of spherical harmonics, $Y_{lm}(\hat{n})$, as $ T(\hat{n}) = \sum_{lm} a_{lm} Y_{lm} (\hat{n})$,
where $\hat{n}$ denotes position coordinates on the sphere, and $a_{lm}$ are spherical harmonic
coefficients. Then, the two point correlation function of the CMB field
$C(\hat{n}_1,\hat{n}_2) = \langle T(\hat{n}_1) T(\hat{n}_2) \rangle$
can be fully described in terms of BipoSH coefficients as~\cite{Hajian03,Basak06}
\be
A^{LM}_{l_1 l_2} = \int d\hat{n}_1 d\hat{n}_2 \, C(\hat{n}_1,\hat{n}_2) \, \{Y(\hat{n}_1) \otimes Y(\hat{n}_2)\}^{LM^*}_{l_1 l_2} \,,
\ee
whose diagonal elements are the much familiar angular power spectrum i.e., $A^{00}_{l_1 l_2} \propto C_{l_1} \delta_{l_1 l_2}$.
Thus BipoSH coefficients capture all possible correlations between various modes of CMB anisotropies. They
are related to the spherical harmonic coefficients of CMB field as
\be
A^{LM}_{l_1 l_2} = \sum_{m_1 m_2} a_{l_1 m_1} a_{l_2 m_2} C^{LM}_{l_1 m_1 l_2 m_2}
\label{eq:biposhdef}
\ee
where $C^{LM}_{l_1 m_1 l_2 m_2}$ are the Clebsch-Gordan coefficients whose indices are related by
$|l_1-l_2| \leq L \leq l_1+l_2$ and $M=m_1+m_2$~\cite{varshalovich}.

\section{BipoSH estimator for recovering SI violating signals in CMB}

\subsection{The estimator}
Here we briefly review some relevant details and the estimator itself, for the recovery
of SI violating signals underlying an observed CMB map. For further details the reader
is directed to consult the primary article~\cite{msfpaper}.

The effect of Doppler boosting on CMB temperature anisotropies~\cite{doppler} can be found to be given by
$T(\hat{n}) = \mathcal{T}_0 \vec{\beta} \cdot \hat{n} + \Theta \left(\hat{n}-\nabla(\vec{\beta}\cdot \hat{n})\right) \left(1+b_\nu\vec{\beta} \cdot \hat{n} \right)$~\cite{plk13dop}, where $\mathcal{T}_0=2.726$~Kelvin is the mean
CMB sky temperature, $b_\nu=(\nu/\nu_0)\coth(\nu/2\nu_0)-1$ is a frequency dependent factor
($\nu_0=k_B \mathcal{T}_0/h \approx 57$~GHz), and $T(\hat{n})$ and $ \Theta(\hat{n})$ are the
boosted and unboosted temperature anisotropies.

The BipoSH coefficients of a full-sky Doppler boosted CMB map are given by
\begin{equation}
A^{LM}_{l_1 l_2} = \left(A^{LM}_{l_1 l_2}\right)_{ub.cmb} + \beta_{LM}H^{L}_{l_1 l_2}\,,
\label{eq:fs-dop-biposh}
\end{equation}
where $\left(A^{LM}_{l_1 l_2}\right)_{ub.cmb}$ are \emph{unboosted} CMB anisotropies (that can have
anisotropic noise), $\beta_{LM}$ are the spherical harmonic coefficients of the Doppler field
$\beta(\hat{n})=\vec{\beta}\cdot\hat{n}$, and $H^{L}_{l_1 l_2}$ is the shape function characterizing
the effect of Doppler boost in harmonic space (see appendix for details).
Note that Doppler boosting generates only $L=1$ modes i.e., it induces coupling between $l$ and $l\pm1$ multipoles.

In the presence of mask or partial sky coverage, the BipoSH coefficients of a boosted sky
can be found to be given by,
\begin{equation}
\tilde{A}^{LM}_{l_1l_2} = \left(\tilde{A}^{LM}_{l_1l_2}\right)_{ub.cmb} + \sum_{L'M'} \beta_{L'M'} K^{L'M'}_{LMl_1l_2} \,,
\label{eq:mask-dop-biposh}
\end{equation}
where $\left( \tilde{A}^{LM}_{l_1 l_2} \right)_{ub.cmb}$ denotes masked unboosted sky BipoSH coefficients,
and $K^{L'M'}_{LMl_1l_2}$ is the masked analogue of the full-sky shape function $H^{L}_{l_1 l_2}$
(see appendix for details). We refer to it as modified shape function (MSF).

Masking introduces leakage of intrinsic anisotropic modes $\{L',M'\}$ to observed modes $\{L,M\}$.
This mixing/leakage is captured by the modified shape function.
However, based on the sky fraction and the broadly azimuthal nature of the galactic masks used in CMB analysis,
some simplifications are in order :
\begin{itemize}
\item The anisotropic signal is predominantly retained in the observed mode that is same as the intrinsic mode i.e., $L=L'$.
\item The MSF is diagonal in phase modes i.e., $M=M'$.
\end{itemize}
Under these approximations i.e., the diagonal MSF sufficiently accounting for the loss
of signal due to mask, Eq.~[\ref{eq:mask-dop-biposh}] reduces to
$\tilde{A}^{L M}_{l_1 l_2} = \left(\tilde{A}^{LM}_{l_1 l_2}\right)_{ub.cmb} + \beta_{LM} K^{L M}_{L M l_1 l_2}$.
This can be inverted to define a minimum variance estimator as
\begin{equation}
\hat{\beta}_{LM} = \sum_{l_1 l_2} \hat{w}^{L}_{l_1 l_2} \frac{\hat{\mathcal{A}}^{LM}_{l_1 l_2}}{K^{LM}_{L\,M\,l_1\,l_2}} \,,
\label{opt_est}
\end{equation}
where $\hat{\mathcal{A}}^{LM}_{l_1l_2} = \tilde{A}^{LM}_{l_1l_2} - \langle \tilde{A}^{LM}_{l_1 l_2} \rangle_{ub.cmb}$ are the mask-bias corrected  BipoSH coefficients from data map, and $\hat{w}^{L}_{l_1 l_2}$ are the weights for linearly combining the BipoSH coefficients that minimize the variance of the reconstructed anisotropic signal.
The weights also satisfy the constraint that they all add up to \emph{one} i.e., $\sum_{l_1 l_2} \hat{w}^L_{l_1 l_2}=1$ so that the anisotropic signal being reconstructed is intact (see appendix for details). The bias term due to mask, $\langle \tilde{A}^{LM}_{l_1 l_2} \rangle_{ub.cmb}$, is estimated from unboosted simulations.
We see that the effective weights are $M$ dependent owing to the $M$ dependent shape function in Eq.~[\ref{opt_est}], unlike the full-sky estimator. In deriving the weights, we made another approximation
that :
\begin{itemize}
\item $\langle \hat{\mathcal{A}}^{LM}_{l_1 l_2} \hat{\mathcal{A}}^{LM}_{l'_1 l'_2} \rangle_{ub.cmb} \approx (\hat{\sigma}^{LM}_{l_1 l_2})^2 \delta_{l_1 l'_1} \delta_{l_2 l'_2}$ i.e., the covariance of unboosted BipoSH coefficients is essentially diagonal.
\end{itemize}

We emphasize that though the equations given are for the case of Doppler boosted CMB
anisotropies, the Doppler field can be taken to be
representative of any isotropy violating field that breaks statistical anisotropy
of CMB, with a corresponding shape function (indeed many
of the isotropy violating effects on CMB upto first order take the generic form
of Eq.~[\ref{eq:fs-dop-biposh}]~\cite{orthobiposh}).

\subsection{Results}

To validate our estimator, we generated a set of 1000 boosted CMB skies at \texttt{HEALPix}~\cite{Gorski05}
$N_{side}=2048$ using the \texttt{CoNIGS}~\cite{conigs} algorithm (Code for Non-Isotropic Gaussian Sky),
and convolved with beam of $fwhm=5$~arcmin corresponding to PLANCK's 217~GHz channel~\cite{PlanckBlueBook}.
These have Doppler boost injected with the observed amplitude of $\beta=1.23\times10^{-3}$
in the actual dipole direction $(l,b)=(264^\circ,48^\circ)$, in Galactic coordinates~\cite{l1dipole}.
Further realistic noise maps are generated
to add with the boosted, beam convolved CMB maps using the PLANCK 2015 full mission
anisotropic noise rms, also corresponding to 217~GHz channel.
We used $b_\nu=3$ to correspond to 217~GHz band for the scale dependent factor
in the shape function of Eq.~[\ref{eq:fs-dop-biposh}] (also see Eq.~[\ref{eq:fs-shape}] in appendix).

In Fig.~[\ref{fig:approximations}], we quantitatively provide justifications to all the three approximations
made in arriving at the masked sky estimator given in Eq.~[\ref{opt_est}]. In Fig.~[\ref{fig:mask}],
the PLANCK 2015 common analysis mask used in our present study is shown. It is used after apodizing with
a Gaussian beam of $fwhm=30$~arcmin. In Fig.~[\ref{fig:msf}], the diagonal nature of MSF
in $\{L,M\}$ is shown. The leakage/mixing of modes due to mask between various $M$
and $M'$ modes for $L=L'=1$ (and additionally from $L'=1$ to $L=2$ for $M=M'=0$) are depicted in the figure.
It can be readily seen that the MSF of the off-diagonal elements are much lower in amplitude than
the diagonal elements. Further we also checked the consistency of the relation
$\hat{\mathcal{A}}^{LM}_{l_1l_2} = \beta_{LM} K^{L M}_{L M l_1 l_2}$,
in Fig.~[\ref{fig:biposhconsistency}], where the LHS is computed from Doppler boosted simulations and the
RHS is obtained analytically. We see that our diagonal approximation to MSF taken on the RHS agrees with the
full simulation estimate on LHS. We also see that the primary source of uncertainty in faithfully
recovering the Doppler signal is the real part of $\hat{\mathcal{A}}^{11}_{l l+1}$ which
is weakest among the three components.
Note that the two terms, LHS and RHS, are plotted in WMAP normalization for
even parity modes~\cite{wmap7yranom}, same as in Fig.~[\ref{fig:msf}] for MSF.
Then, the normalized covariances of the mask-bias corrected
BipoSH coefficients from unboosted simulations are shown for real part of $M=M'=0$, and real and imaginary parts
of $M=M'=1$ in Fig.~[\ref{fig:biposhvarM0}], [\ref{fig:biposhvarM1r}] and [\ref{fig:biposhvarM1i}],
respectively for $L=L'=1$. The normalized covariance matrices are defined as $R_{ll'} = C_{ll'}/\sqrt{|C_{ll}||C_{l'l'}|}$
where $C_{ll'} = \langle \hat{\mathcal{A}}^{1M}_{l l+1} \hat{\mathcal{A}}^{1M^*}_{l' l'+1} \rangle_{ub.cmb}$.
We clearly see that the covariance matrices corresponding to the multipole range $l=[2,2000]$
are essentially diagonal, and the off-diagonals are at ten percent level or less compared to the
diagonal terms.

\begin{figure}
\centering
\subfloat[\label{fig:mask}]{\includegraphics[width=0.45\linewidth]{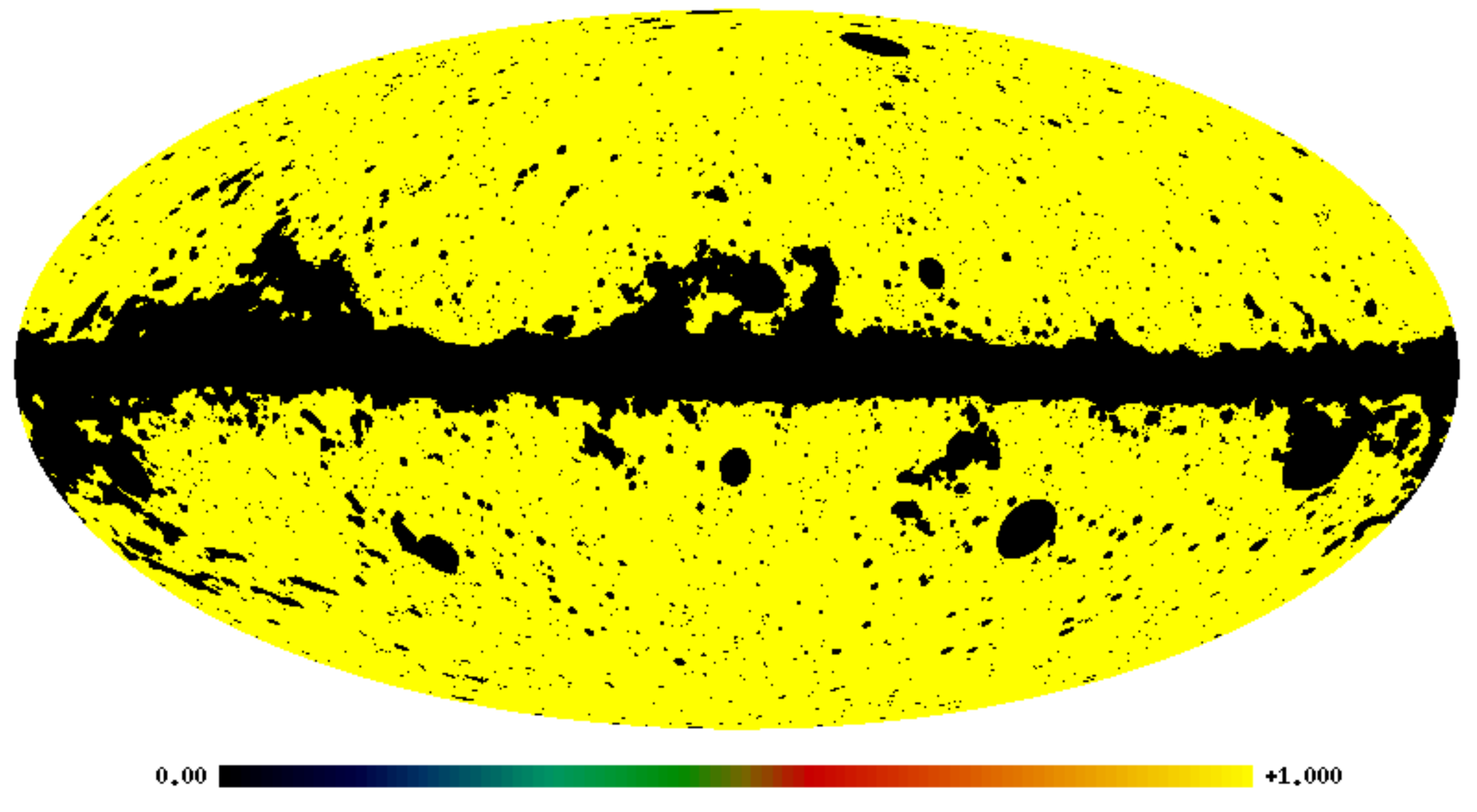}}
\hfill
\subfloat[\label{fig:msf}]{\includegraphics[width=0.54\linewidth]{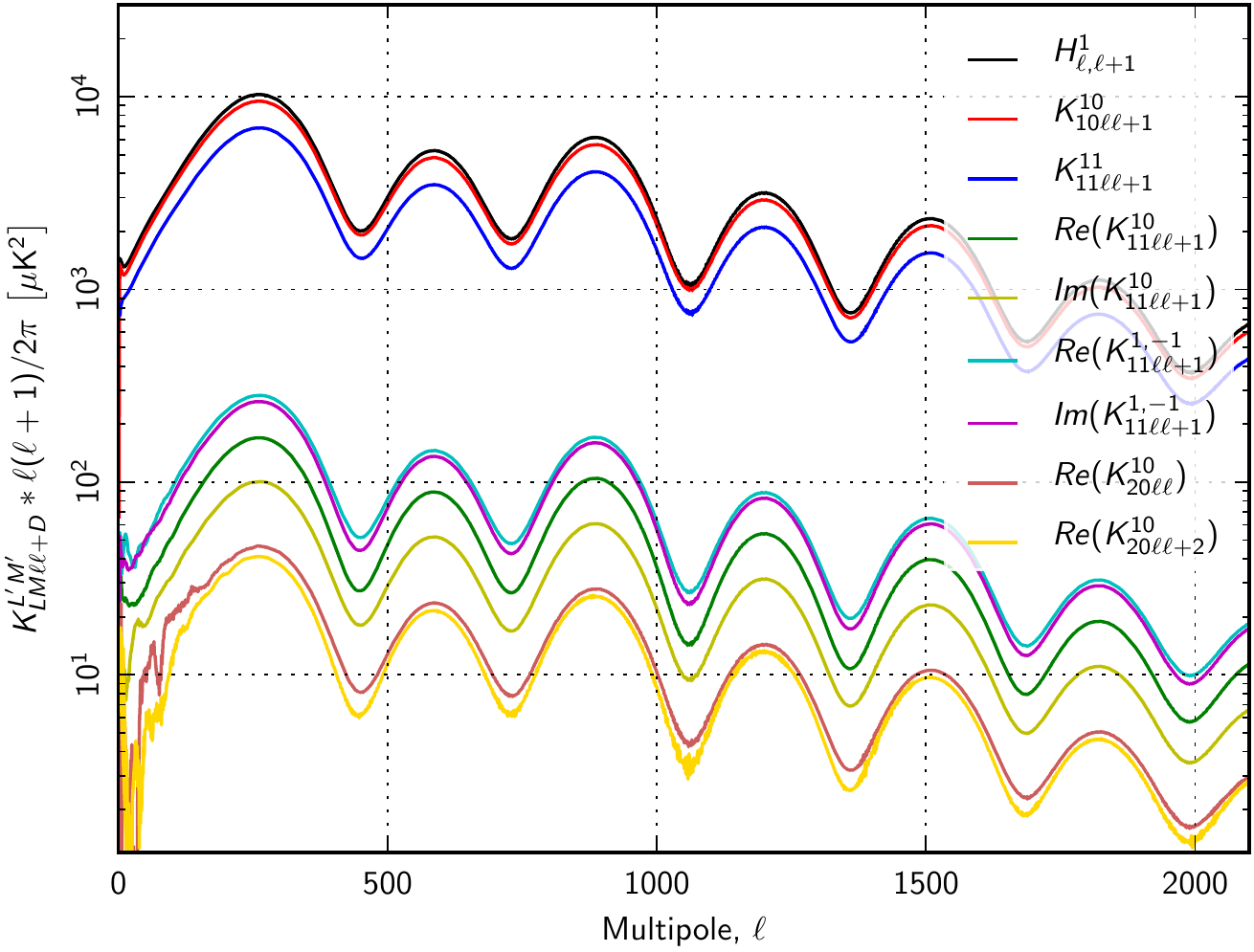}}
\hfill
\subfloat[\label{fig:biposhconsistency}]{\includegraphics[width=0.85\linewidth]{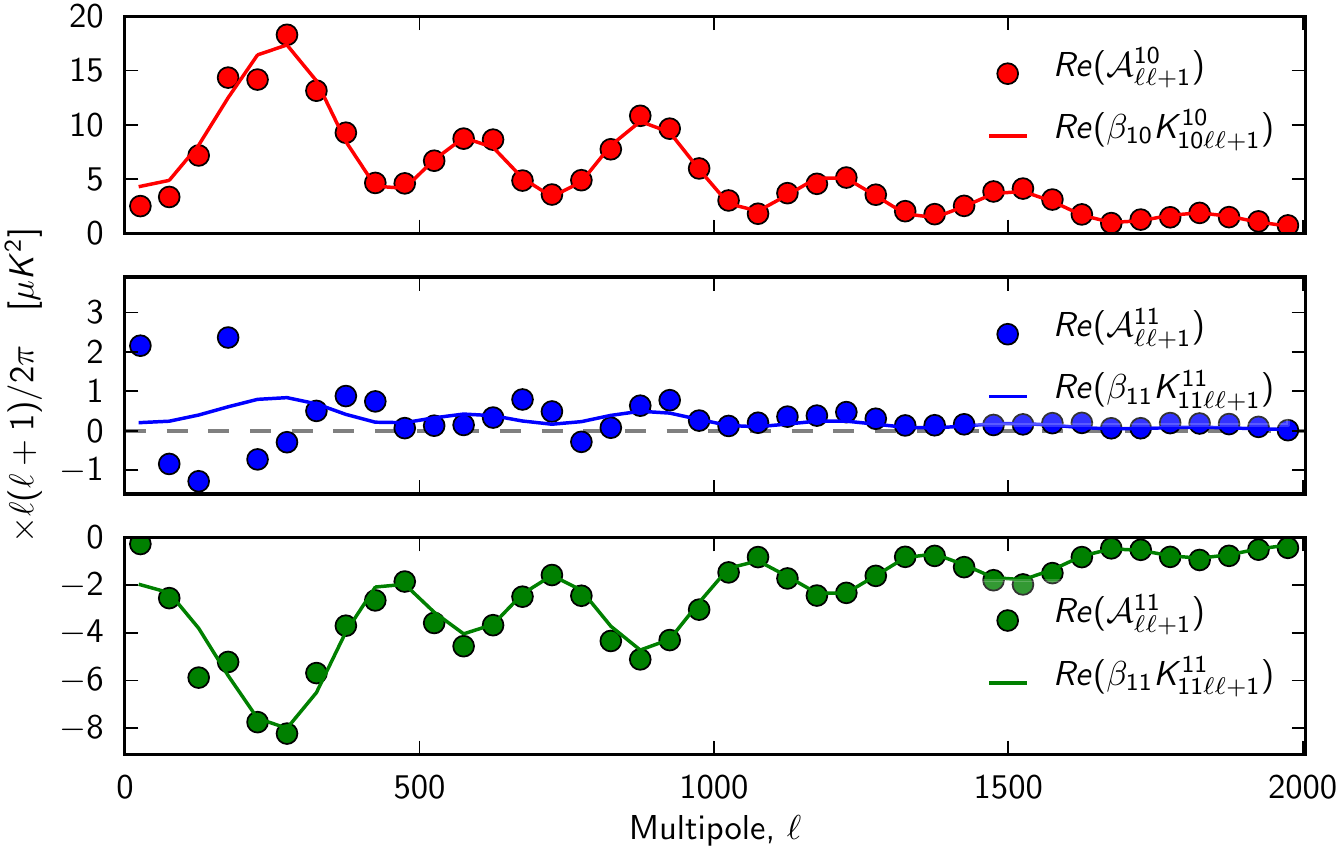}}
\hfill
\subfloat[\label{fig:biposhvarM0}]{\includegraphics[width=0.325\linewidth]{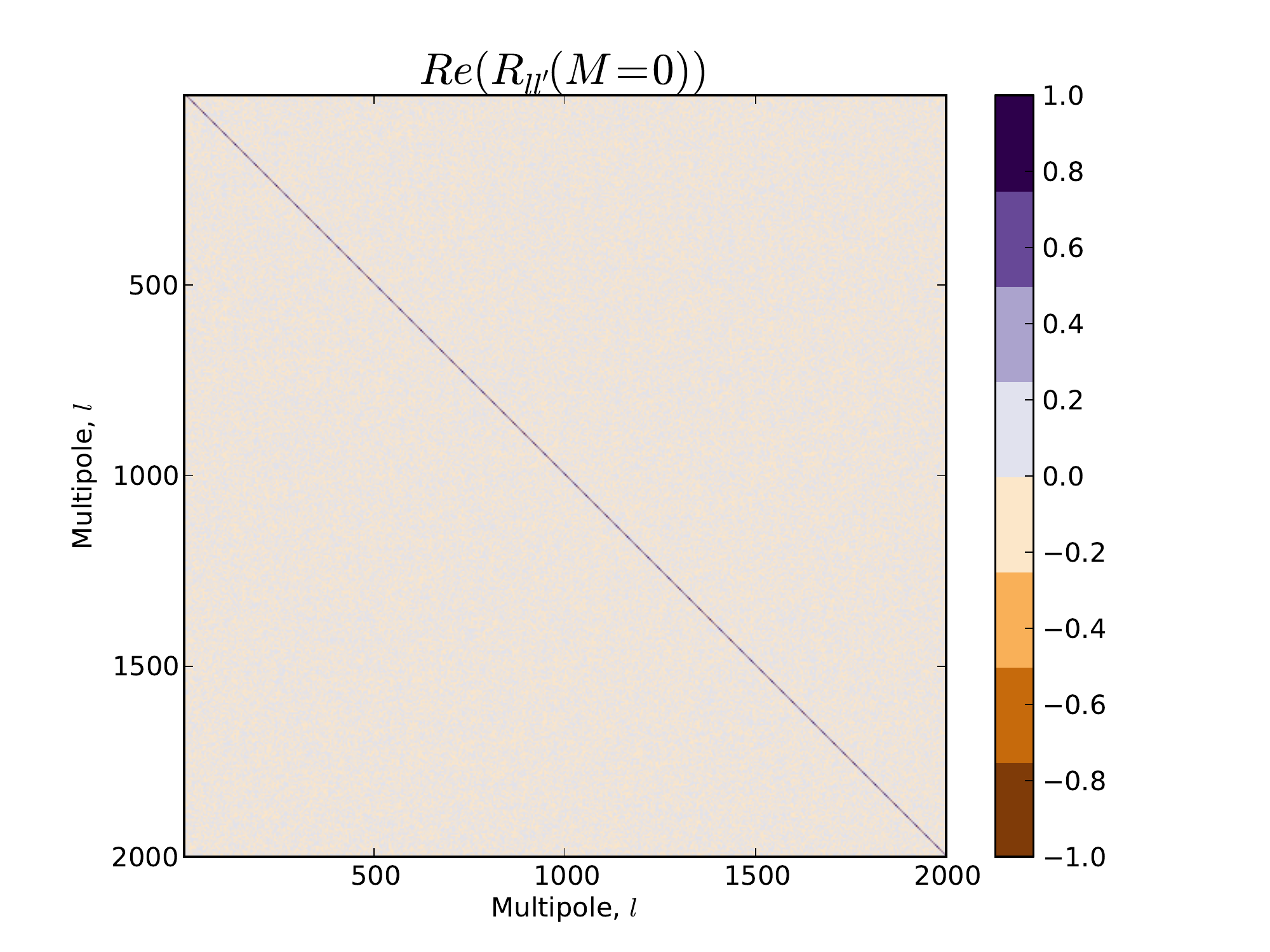}}
\hfill
\subfloat[\label{fig:biposhvarM1r}]{\includegraphics[width=0.325\linewidth]{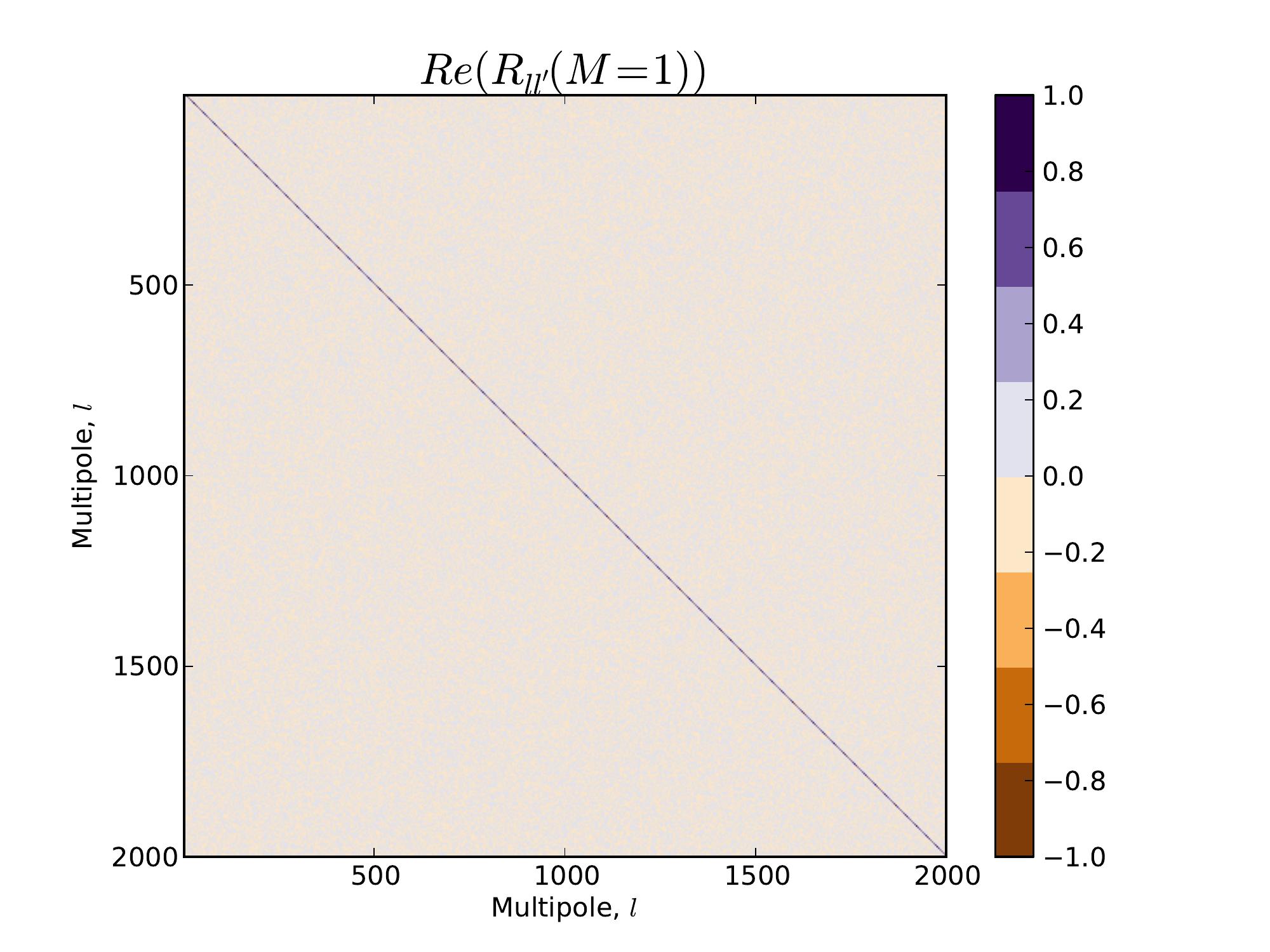}}
\hfill
\subfloat[\label{fig:biposhvarM1i}]{\includegraphics[width=0.325\linewidth]{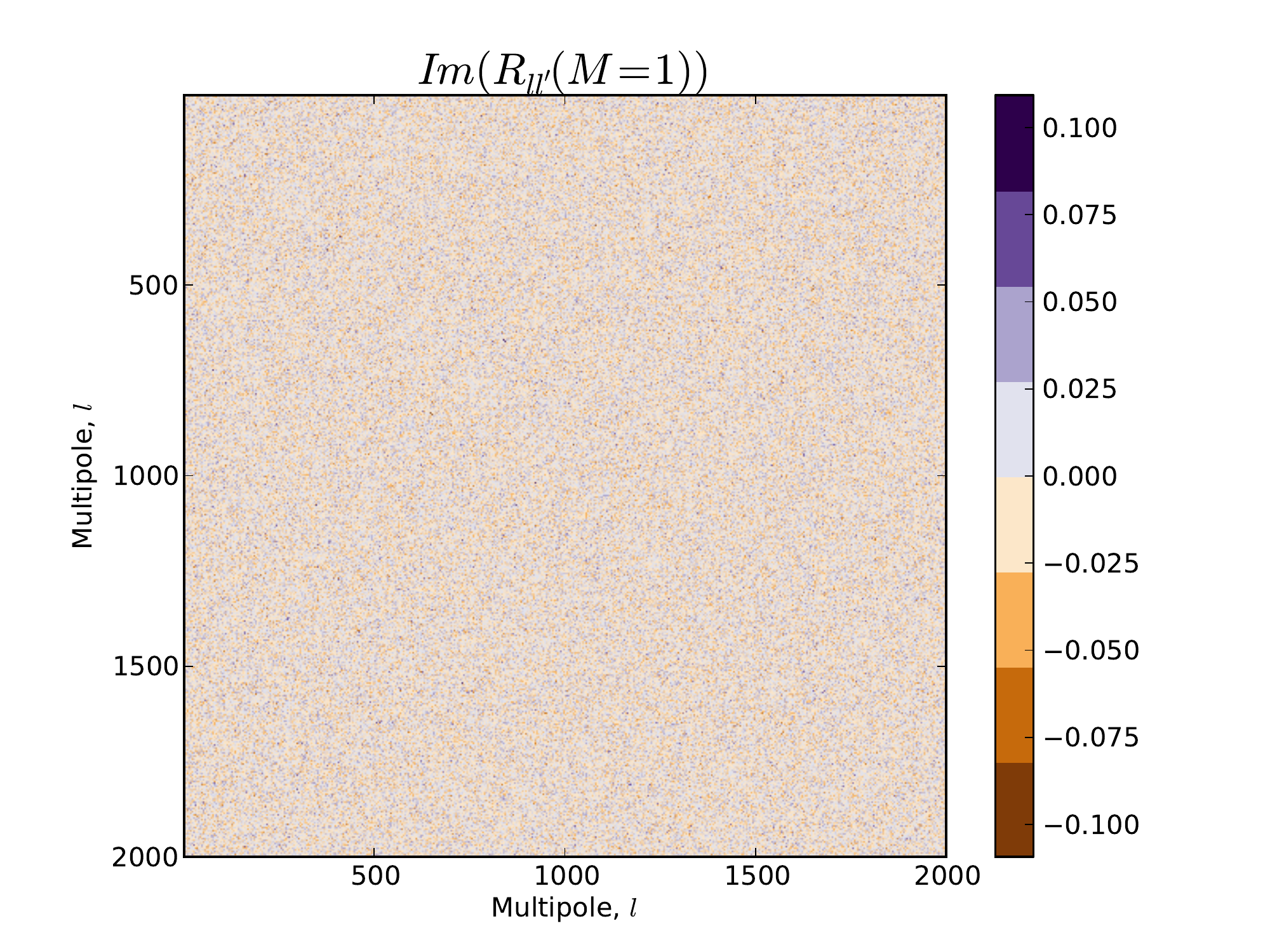}}
\caption{The approximations made in arriving at the masked sky estimator given in Eq.~[\ref{opt_est}]
         are validated in this figure : $(a)$~PLANCK 2015 common analysis mask (with point sources)
         that is used in the present analysis. $(b)$~MSF of the Doppler boost signal for various
         combinations of intrinsic anisotropic modes $\{L',M'\}$ and observed modes $\{L,M\}$.
         $(c)$~Verifying the consistency of the relation
         $\tilde{A}^{L M}_{l_1 l_2} - \langle\tilde{A}^{LM}_{l_1 l_2}\rangle_{ub.cmb} = \beta_{LM} K^{L M}_{L M l_1 l_2}$
         after making the diagonal approximation to MSF. $(d)$, $(e)$ \& $(f)$~- Covariance of the mask-bias
         corrected unboosted BipoSH coefficients i.e.,
         $\langle \hat{\mathcal{A}}^{1M}_{l l+1} \hat{\mathcal{A}}^{1M^*}_{l' l'+1} \rangle_{ub.cmb}$
         corresponding to $L=L'=1$ and real part of $M=M'=0$ ($d$) and real \& imaginary parts of $M=M'=1$
         modes ($e$ \& $f$), respectively.}
\label{fig:approximations}
\end{figure}

\begin{figure}
\centering
\includegraphics[width=0.6\textwidth]{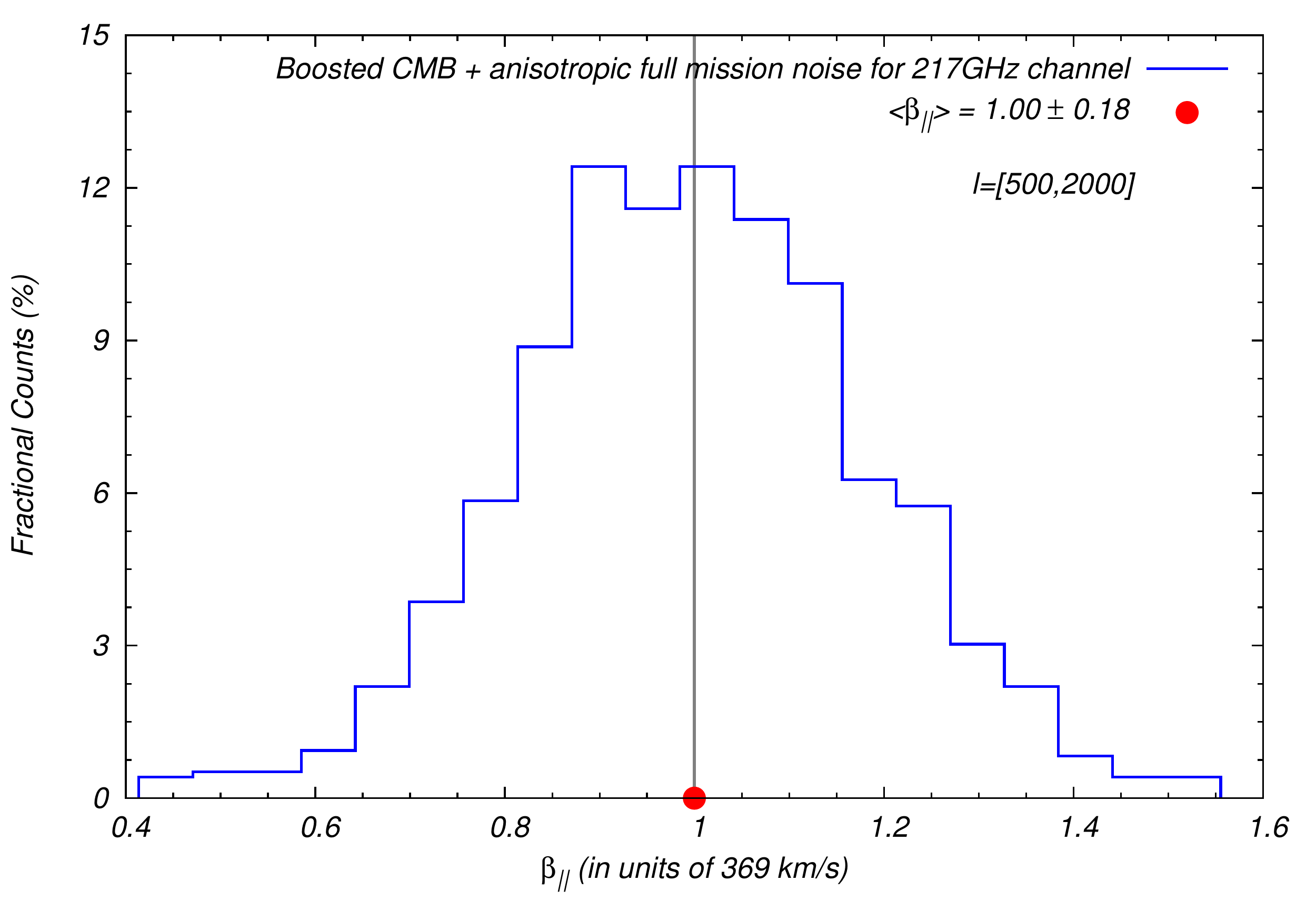}
\caption{Shown here is the histogram plot of
         recovered Doppler signal from 1000 boosted simulations added with anisotropic noise
         realizations corresponding to 2015 full mission data from PLANCK's 217~GHz channel.
         The results are presented in terms of PLANCK 2013 Doppler estimate `$\beta_{\parallel}$' ie.,
         projection of the recovered Doppler amplitude along the known CMB dipole direction
         $(l,b)=(264^\circ,48^\circ)$, in units of the known Doppler amplitude
         $\beta=1.23\times10^{-3}$.}
\label{fig:projbeta}
\end{figure}

The recovered Doppler signal from the simulations is cast in terms of the $\beta_\parallel$
estimate used in PLANCK 2013 Doppler paper~\cite{plk13dop}.
$\beta_\parallel=(\beta/\beta_0)\hat{\beta}\cdot\hat{\beta}_0$ is the recovered Doppler vector,
$\vec{\beta}=(\beta,\hat{\beta})$, projected along the known dipole direction ($\hat{\beta}_0$)
in units of the known Doppler amplitude ($\beta_0 = 1.23 \times 10^{-3}$).
Note that the amplitude is computed from the recovered Doppler power, $\beta_1=\sum_M |\beta_{1M}|^2/3$,
after correcting for the reconstruction noise bias, $\beta^N_1$, which is estimated by applying
the partial sky estimator to unboosted simulations. The Doppler amplitude and power are related by
$\beta = 1.5\sqrt{\beta_1 - \beta^N_1}$.
However, the direction is estimated directly from the recovered Doppler map.
The results are shown in Fig.~[\ref{fig:projbeta}].
We obtain $\beta_\parallel=1.00\pm0.18$, using the multipole window $l=[500,2000]$,
which has a slightly lower error on face value compared to the PLANCK 2013 Doppler estimate
from the 217~GHz observed map i.e., $\beta_\parallel=0.75\pm0.19$.

\section*{Acknowledgments}
We thank Suvodip Mukherjee for generating Doppler boost simulations using \texttt{CoNIGS}~\cite{conigs} algorithm.
We acknowledge the use of ESA's PLANCK data from public release 2, and \texttt{HEALPix}~\cite{Gorski05}
software package for CMB analysis.

\section*{Appendix}
The fullsky Doppler boost shape function is given by
\begin{subequations}
\begin{align}
H^{L}_{l_1 l_2} &= b_\nu \left(G^{L}_{l_1 l_2}\right)_{mod} - \left(G^{L}_{l_1 l_2}\right)_{abr} \,, \\
\left(G^{L}_{l_1 l_2}\right)_{mod} &= \frac{C_{l_1} + C_{l_2}}{\sqrt{4\pi}}
\frac{\Pi_{l_1}\Pi_{l_2}}{\Pi_{L}} C^{L 0}_{l_1 0 l_2 0} \,, \\
\left(G^{L}_{l_1 l_2}\right)_{abr} &= \frac{\left[ C_{l_1} F(l_1,L,l_2) + C_{l_2} F(l_2,L,l_1) \right] }{\sqrt{4\pi}} \frac{\Pi_{l_1}\Pi_{l_2}}{\Pi_{L}} C^{L 0}_{l_1 0 l_2 0} \,, \\
F(l_1,L,l_2) &= \frac{l_1(l_1 + 1) + L(L + 1) - l_2(l_2 + 1)}{2} \,,
\end{align}
\label{eq:fs-shape}
\end{subequations}
where $\left(G^{L}_{l_1 l_2}\right)_{mod}$ and $\left(G^{L}_{l_1 l_2}\right)_{abr}$ denote shape functions
corresponding to the modulation and the aberration parts of the Doppler boost respectively, $\Pi_l=\sqrt{2l+1}$,
and $C_l$ is the fiducial power spectrum, also used to generate isotropic and Doppler boosted CMB realizations.

The modified shape function (MSF) due to mask or availability of partial sky is given by
\begin{eqnarray}
K^{L'M'}_{LMl_1l_2} = \sum_{l_3 l_4} \Pi_{L'} H^{L'}_{l_3 l_4} \frac{\Pi_{l_3} \Pi_{l_4}}{\sqrt{4\pi}} \sum_{l_5 l_6} \frac{\Pi_{l_5} \Pi_{l_6}}{\sqrt{4\pi}} C^{l_1 0}_{l_3 0 l_5 0}
 C^{l_2 0}_{l_4 0 l_6 0} \sum_{J K}	\left\{
									\begin{array}{c c c}
									L  & l_1 & l_2 \\
									L' & l_3 & l_4 \\
									J  & l_5 & l_6
									\end{array}
									\right\}
									\Pi_{J} W^{J K}_{l_5 l_6}  C^{L M}_{L' M' J K}\,,
\label{eq:mixingkernel}
\end{eqnarray}
where the quantity $\left\{ \right\}_{3\times3}$ is the Wigner $9j-$symbol, and $W^{J K}_{l_5 l_6}$ are
the BipoSH of mask used, which are obtained from Eq.~[\ref{eq:biposhdef}] using the spherical harmonic
coefficients of mask. From the Clebsch-Gordon coefficient $C^{L M}_{L' M' J K}$ in the above equation,
we can see that the intrinsic anisotropic modes $\{L',M'\}$ mixes with the mask modes $\{J,K\}$ giving
rise to the observed modes $\{L,M\}$.

The weights $ \hat{w}^{L}_{l_1 l_2}$ which minimize the variance are given by,
\begin{equation}
\hat{w}^{L}_{l_1 l_2} = \frac{1}{\sum_{M} \left(\frac{\hat{\sigma}^{LM}_{l_1 l_2}}{K^{LM}_{L M l_1 l_2}}\right)^2}
\left[
\sum_{l'_1 l'_2} \frac{1}{\sum_{M} \left(\frac{\hat{\sigma}^{LM}_{l'_1 l'_2}}{K^{LM}_{L M' l'_1 l'_2}}\right)^2}
\right]^{-1} \,,
\end{equation}
where,
\begin{equation}
 \left(\hat{\sigma}^{LM}_{l_1 l_2}\right)^2 = \langle |\tilde{A}^{LM}_{l_1 l_2}|^2 \rangle_{ub.cmb} -
     |\langle \tilde{A}^{LM}_{l_1 l_2} \rangle_{ub.cmb}|^2 \,,
\end{equation}
is the variance of unboosted map's BipoSH coefficients for an $\{L,M\}$ mode, that is estimated
from simulations along with the mean mask-bias, $\langle \tilde{A}^{LM}_{l_1 l_2} \rangle_{ub.cmb}$.


\section*{References}

\begin{thebibliography}{99}
\bibitem{wmap7yranom}
 Bennett C. et al., 2011, ApJS, 192, 17
\bibitem{plk13is}
 Ade P. A. R. et al., Planck 2013 results - XXIII, 2014, A\&A, 571, A23
\bibitem{plk15is}
 Ade P. A. R. et al., Planck 2015 results - XVI, 2015 [arXiv:1506.07135]

\bibitem{Hajian03}
 Hajian A., and Souradeep T., 2003, ApJ, 597, L5
\bibitem{Basak06}
 Basak S., Hajian A., and Souradeep T., 2006, Phys. Rev. D, 74, 021301

%
\bibitem{varshalovich}
Varshalovich D. A., Moskalev A. N., and Khersonskii V. K., 1988, \emph{Quantum Theory of Angular Momentum}, World Scientific.

\bibitem{msfpaper}
 Aluri P. K., Pant N., Rotti A., and Souradeep T., 2015 [arXiv:1506.00550]

\bibitem{doppler}
 Challinor A., and van Leeuwen F., 2002, Phys. Rev. D, 65, 103001;
 Burles S., and Rappaport S., 2006, ApJ, 641, L1;
 Kosowsky A., and Kahniashvili T., 2011, Phys. Rev. Lett., 106, 191301;
 Amendola L. et. al., 2011, JCAP, 07, 027;
 Chluba J., 2011, MNRAS, 415, 3227;
 Notari A., and Quartin M., 2012, JCAP, 02, 026

\bibitem{plk13dop}
 Aghanim N. et al., Planck 2013 results - XXVII, 2014, A\&A, 571, A27

\bibitem{orthobiposh}
 Kumar S et~al., 2015, Phys. Rev. D, 91, 043501

\bibitem{Gorski05}
 Gorski K. M. et al., 2005, ApJ, 622, 759

\bibitem{conigs}
 Mukherjee S., and Souradeep T., 2014, Phys. Rev. D, 89, 063013

\bibitem{PlanckBlueBook}
Planck `Blue Book' : The Scientific Programme of Planck, Planck Collaboration, 2005, ESA-SCI(2005)01
%
\bibitem{l1dipole}
Kogut A. et. al., 1993, ApJ, 419, 1;
Fixsen D. et. al., 1996, ApJ, 473, 576;
Hinshaw G. et. al., 2009, ApJS, 180, 225
%
\end{thebibliography}
\end{document}